\documentclass{sf2a-conf2025}
\usepackage{graphicx}
\usepackage{hyperref}
\usepackage[]{natbib}  
\usepackage{epstopdf}

\def\BibTeX{{\rm B\kern-.05em{\sc i\kern-.025em b}\kern-.08em
    T\kern-.1667em\lower.7ex\hbox{E}\kern-.125emX}}
\bibpunct{(}{)}{;}{a}{}{,}  %%%%%%%%%%%%%  A&A bibliography style
\begin{document}

\TitreGlobal{SF2A 2025}

%%-----------------------------------------------------------------
%%      the top matter
%%

\title{SF2A Environmental Transition Commission:\\ Summary of the 2025 workshop}

\runningtitle{Review SF2A-2025 S13}

\author{F.~Cantalloube}\address{Univ. Grenoble Alpes, CNRS, IPAG, F-38000 Grenoble, France}
%% SOC-S13
\author{J.~Berat}\address{Laboratoire de Physique de l’\'Ecole Normale Sup\'erieure, ENS, Universit\'e PSL, CNRS, Sorbonne Universit\'e, Universit\'e de Paris, F-75005 Paris, France}
\author{N.~Fargette}\address{IRAP, Université de Toulouse, CNRS, CNES, UT3-PS, Toulouse, France}
\author{P.~Larue$^1$}
\author{N.~Pourré$^1$}
\author{S.~Anderson}\address{Aix-Marseille Univ, CNRS, CNES, LAM, Marseille, France}
%\author{A.~Santerne$^{1,4}$}
\author{J.~Milli$^1$}
\author{J.-F.~Gonzalez}\address{Universite Claude Bernard Lyon 1, CRAL UMR5574, ENS de Lyon, CNRS, Villeurbanne 69622, France}
\author{O.~Berné$^3$}
\author{D.~Barret$^3$}
%% Intervenant·es S13
\author{A.~Mouinié$^3$}
\author{C.~Picard}\address{Observatoire de Paris - PSL, 61 avenue de l’Observatoire, 75014 Paris, France}
\author{K.~Dassas}\address{CESBIO, Université de Toulouse, 31000 Toulouse, France}
\author{J.~Knödlseder$^3$}
\author{D.~Redon}\address{Pouvoirs, Histoire, Esclavage, Environnement, Atlantique, Caraïbe / UMR-CNRS 8053, LABEX CEBA}

%% Keep this line, even if the page will be settled afterwards.
%\setcounter{page}{}

%%-----------------------------------------------------------------

\maketitle

%%-----------------------------------------------------------------
%%        The abstract
%% 
%%  Warning!  within the abstract:
%%  - do not use macros. 
%%  - do not use commands like: \cite, \citet, \citep ... etc.

\begin{abstract}
During its annual conference in 2025, the French Society of Astronomy \& Astrophysics (SF2A) hosted, for the fifth time, a special session dedicated to discussing the environmental transition within the French A\&A research community. 
During the 2025 workshop, the goal was to review four contemporary topics within the context of environmental transition actions and discussions: (1) institutional actions, (2) the early-career researchers singularity, (3) research infrastructures and tools, and (4) the geopolitical conditions under which A\&A research remains possible. 
The workshop concluded with a round-table discussion that brought together the various speakers so that every participant could express their ideas.  
\end{abstract}

%% Insert the keywords (to appear in the ADS indexing)
%% Keywords must be separated by a comma

%%-----------------------------------------------------------------

\section{Introduction}
%%---------------------
Since 2021, the Environmental Transition workshops during the SF2A annual conference, have been bringing together the community of Astronomy \& Astrophysics researchers to reflect on existing or potential means of action to adapt our professions to the accelerating environmental crisis. This publication summarizes the main messages from each contributed talk, pointing towards the corresponding publication. It also reports the main topics discussed during the 30~min round-table. Around eighty people participated in this afternoon session.

%%---------------------  
\section{Summary of actions taken at an OSU level since 2017 : the case of OMP}
The first introductory talk was given by A.~Mouinié, transition officer for the Observatory of Universe Science (OSU) Observatoire Midi-Pyrénées (OMP). The speaker, A.~Mouinié, took this position after being in charge of the transition at IRAP, the astrophysics institute of Toulouse. The benefit of working at the larger level of an OSU (an organization made up of several research institutes working in the natural sciences) is that it unites the actions undertaken by the different labs, yet remaining local. This strategic scale makes it possible to increase the animating forces, gives greater weight in discussions with the various tutelles, provides more means, and encourages the sharing of experiences and resources. It also takes advantage of the different fields of study represented by the various labs that make up the OSU, especially the climate and environmental sciences. %As an example, operations such as MT180 (OSUG), Villarceaux agreements (IPSL) or Festi'Climat (OMP) were initiated by OSUs. % ADD LINKS  echelle stratégique `

The French National Centre for Scientific Research (CNRS) published in 2025 its \emph{Sustainable Development and Social Responsibility} scheme\footnote{\url{https://www.cnrs.fr/sites/default/files/news/2025-03/_Schema_Directeur_DDRS_CNRS.pdf}.}. In this document, the Universe Science Institute of CNRS (INSU) requires the directors of each institutes to commit to a greenhouse gas emissions (GHG) reduction trajectory of 6\% per year. To follow this, a tool to quantify the GHG emissions of research institutes is under development at CNRS, based on the \emph{GES-1point5} tool developed within the \emph{Labos1point5} research group \citep{mariette2022}. Such initiative was supplemented by the University of Toulouse and the French National Research Institute for Sustainable Development (IRD), who established a strategic plan to reduce their environmental footprint.

In 2017, the OMP environmental transition commission started its activity by creating the workshop Ecology-Environment-Transition (EET). This included initiative such as training members to estimate GHG footprints, organizing three editions of the \emph{Festi’Climat}, and settling the role of the commission. Each institute constituting the OMP has two representatives taking part to the the commission. One notable action is a follow-up of the GHG footprint estimation of each institutes across years (without accounting for the research infrastructures). A comparison of these estimates, normalized on the numbers of employees, was published in \cite{marc2024omp}. %Such comparison is relevant for relatively large OSU.

The main action is the establishment of a road map with an adapted strategy to reduce the environmental impacts of the research activities. The road map is established along 5 different poles: training, instrumentation, data and numerical, structural operation and national observation services (SNO). The commission of OMP aims at ensuring consistence in the transition objectives of the OMP, setting common goals and guidelines, contributing to the follow-up and evaluation of the transition process and proposing concrete solutions to reduce the GHG emissions towards changing and reforming research beyond individual interests.

%Questions: liens acec les SHS et Atecopol par ex

%%-------------------------
\section{Early-career researchers' facing climate and ecological emergency}
The second presentation, given by J.~Berat, highlighted and discussed the specificity of early-career researchers (ECR), suffering from contradictory injunctions and academic pressure. Numerous publications and reports pinpoint the greater vulnerability of PhD candidates to mental health issues. For instance, in 2014, the UC Berkeley \emph{Graduate Student Happiness \& Well-Being Report}\footnote{https://ga.berkeley.edu/wp-content/uploads/wellbeingreport.pdf} revealed that 47\% of undergraduate students reported experiencing signs of depression. In recent years, to this already stressful situation, this population has faced additional stress from climate and ecological emergencies.

In such situation, to avoid the strong feeling of isolation and the difficulty to act at the individual level, one solution is to form discussion groups and confrontation spaces with colleagues of any status. However, the topic remains difficult to bear and express within the community. There is also in general little recognition for interdisciplinary studies and discussions. According to \cite{dablander2024}, once one is willing to commit, the main hindrance to commitment is sorted into two main barriers: practical (lack of skills, fear of loosing credibility, isolation, lack of time and/or opportunities) and intellectual (not the role of scientist, not feeling appropriate, being also part of the problem). Some specific elements for early-career researchers exposed by the speaker are the following: (1) the recruitment criteria are not in line with a form of commitment \cite{gardner2021}, (2) job insecurity and instability is hampering possibilities of long-term commitment, (3) sometimes the lack of support from supervisors, recruiters, or colleagues prevents from accepting commitment, (4) a loss of purpose in one's work may paralyze the person in both their work and commitment, and (5) a fundamental questioning of the ambiguous role of science and scientists\footnote{For instance the following IFOP report highlights that, in France, science outcomes are perceived as less positive compared to 50 years ago: \url{https://www.ifop.com/publication/le-rapport-des-francais-a-la-science-et-au-progres-scientifique/}} may provoke important cognitive dissonance.

As an answer to all these thoughts, a collective of early-career researcher wrote a platform in a French newspaper Lib\'eration: "For a socio-ecological metamorphosis of research"\footnote{\url{https://www.liberation.fr/idees-et-debats/tribunes/pour-une-metamorphose-socioecologique-de-la-recherche-20250106_6NH6KQSM3FCXLHRBAEQAZNL7S4/}}, calling for a redefinition of the profession of researcher that gathered more than 1500 signatures.

%%---------------------------------
 \section{Research infrastructures for astronomy \& astrophysics}

\subsection{Digital carbon footprint at the Paris observatory}
As part of the \textit{Sustainable Development and Social and Environmental Responsibility Master Plan} of the Paris Sciences et Lettres university, 
in collaboration with the \emph{EcoInfo} CNRS research group, the presented approach aimed at contributing, improving and disseminating the knowledge on the environmental impacts of digital technologies. The methodology consisted of an analysis of equipment, its use, and the use of data centers (but without taking AI into account). Emissions mainly originate from the manufacturing phase, which is even more evident for workstations. Digital technologies accounts for 7\% of the institution's total emissions ($0.36~tCO2_{eq}/$pers). It is necessary to anticipate the data storage needs of future projects, as for the Square Kilometre Array (SKA), while continuing to assess the environmental impact of these digital tools and making progress in optimizing their energy consumption. 
The results presented are published in \cite{dassas2025}.

\subsection{Environmental impacts of research infrastructures}
According to previous studies, about 60\% of the annual carbon footprint of a professional astronomer in France ($20~tCO2_{eq}$) are coming from research infrastructures \footnote{\url{https://prospective-aa.sciencesconf.org/data/rapport_synthese_groupeI2_transition_carbone_ecologique.pdf}}. To go further, the environmental impacts of research infrastructures were assessed through Life Cycle Assessement in two cases: a ground-based facility  \cite[the Mid-Sized Cherenkov Telescope Array  at La Palma observatory,][]{dossantos2024} and a space-based facility \citep[the X-IFU instrument of the space mission Athena,][]{barret2024}. From this analysis, solutions to decrease the environmental footprint are proposed on the ground-based facility, leading to a decrease of about 30\%. Knowing this and considering the growth in the number of astronomy infrastructures and their carbon footprint, the authors assessed the trajectory required to meet the objective of the Paris Agreement \cite{knodlseder2024}: the community would need to reduce the number of infrastructures by 3\% per year \emph{and} doubling the decarbonization of existing facilities (note that the environmental impact of dismantling a telescope has not been assessed yet). 
The results presented are published in \cite{knodlseder2025}.
%Impact exprimé en point: 1 pt = impact d'un humain sur une année. 

\subsection{An environmental history of the Kourou spaceport}
This contribution was presented by D.~Redon, PhD candidate in History, on a project entitled \textit{Environmental history of the Guiana Space Center spaceport (CSG), from 1962 to 2025}. Working with the archives of the French national space agency (CNES), one objective of this work is to pay tribute to the native population of the spaceport region located northwest of Kourou in French Guiana. Indeed, the story of the Guiana Space Center, dubbed \textit{Europe's Spaceport}, in the 60's is set in the era of scientific colonialism. Integrated into the French Republic in 1946, and first considered as a \emph{terra nullius}, the area got selected as a space launch site in 1964 after the \emph{Evian agreements} to take over the Hammaguir space base in Algeria, prevailing over Algeria's independence. From 1964 to 1992 the constantly expanding CGS received massive infrastructure investments, marked by expropriation with relocation of the local population in the new segregated town created for the space center, and land anthropization with rapid groundwork including mountain leveling and mining activities \citep{Chambaz2021}. For instance, to meet energy needs, the Petit Saut dam was built in 1992, engulfing $365~\mathrm{km^2}$ of primal forest. It resulted into a control of lands and customs by considerably transforming the population lifestyle and the ecological balance of the CSG area. In parallel, despite the preventive archaeological discoveries, the layout of nature reserves and the presence of debates questioning the validity of this undertaking, the site remains completely mired into the military-industrial complex typical of the 60s. In operation since 1968, the last three decades has seen the establishment of ecological and biodiversity management plans for the CSG region, a technsolutionist scheme mainly performing green-washing \citep{Ferdinand2019}. 
%These results are published in \cite{redon2025}.
%subsequent to building the CGS
%published in mainstream media
%% privilegiant les population non-locales

\subsection{Business trips: shifting from plane to train}
This presentation was given by N.~Fargette on behalf of a group of 6 people who discussed and shared their business trip logbook. 
The author highlighted several steps required when one is willing to take the train instead of the plane: (1) getting the travel authorization, (2) obtaining the budget, (3) optimize the journey's logistic, (4) book the journey, (5) dealing with the contingency during the actual journey. Each step are making up for their own difficulty and require a non-negligible yet invisible amount of work. Despite several positive aspects, the shift to train ride often include to trim on time for research or private life, and is often tedious. On top of incentives to take the train, along with an adapted travel policy, the group proposed a number of suggestions: (i) rethinking conference time to make it possible to travel by train, (ii) recognize and support -either financially or with days off- efforts made to use the train, and (iii) reduce the growing consumption of conference by pooling, delegating and combining the business trips. 
The results presented are published in \cite{Fargette2025}.

%%-------------------------
\section{Research activities facing fascism}
The 6-month-old government of the United States of America led by elected president D.~Trump, has launched a frontal attack on the functioning of public scientific research. In a very mediatic way, academic freedom has been severely affected by the use of ideological censorship, data suppression, and brutal budget cuts, laying off hundreds of scientists \citep{goldman2025}. In addition, numerous student visas were revoked, and several propalestinian supporters on campuses were arbitrarily arrested. 
In response to this, the \emph{Stand Up For Science} movement has emerged in the US, and support followed in France. Following this, several initiatives to welcome scientists at risk were decided, such as the \emph{Safe Place for Science} program, or the \emph{Choose France for Science Program}. It is however unclear how these programs are truly able to protect academic freedom efficiently. Yet, the United States has a global influence in all areas of the Western world, so decisions taken by the US-government have a direct impact on European policies. In astronomy, concerns have been raised because there are several important shared databases and infrastructures, particularly in terms of instrumentation projects, but also numerous academic exchanges (students, postdoctoral researchers, and visiting scholars).
%In astronomy, three main effects are playing a role to the concern raised by the recent US policy: first, the cultural context is similar, and Europeans are steeped in USA culture, which makes identification easier; second, there are several important shared databases and infrastructures, particularly in terms of instrumentation projects, but also numerous academic exchanges (students, postdoctoral researchers, or visitors); third, we share and use the same tools, especially communication tools (the English language, to name but one). 
%This raised high concern in France and neighboring countries. %supported by the European commissionsince the invasion of

Yet, many other countries are facing comparable obscurantist offensives, such as Brasil \citep{carlotto2022bolsonaro} and Argentina \citep{dupret2016milei}. Similarly, since the invasion of Ukraine by Russia in February 2022, Ukrainian researchers have had no choice but to leave their homeland or remain as reservists \citep{unesco2024}. In response, the main French institutes condemned the Russian invasion and suspended all cooperation with that country. 
Even more, the United Nations (UN) have recognized that a scholaticide has been committed in the Gaza strip\footnote{https://www.ohchr.org/fr/press-releases/2024/04/un-experts-deeply-concerned-over-scholasticide-gaza}. This time, the supporters in France were evacuated by law enforcement. 

It turns out that the situation in France is not as idyllic as it might want to appear. %with the above-mentioned initiatives. 
Indeed, in 2024 a governmental decree cut the budget allocated to public and private research by 900~millions euros, about $0.3\%$ of the total budget, with the cuts distributed unequally depending on the theme. In 2025, the cut grew of an additional 1.1~billion euros, to which was added a parliamentary cut of 1.5~billion euros. The highly controversial \textit{Loi de programmation de la recherche (LPR)}, enacted in December 2020, undermines higher education and research in France by increasing job insecurity, promoting competitive funding over stable support, deepening institutional inequalities, and prioritizing bureaucratic performance metrics over academic freedom and long-term scientific inquiry. Under-investment in universities also affect the quality of life of staff and students (20\% of students live below the poverty line). The law was pushed through despite significant opposition from universities, unions, and researchers, and large-scale protests and petitions were largely ignored.
% as for instance the registration fees increase steadily
%This decrease in long-term funding is offset by an increase in multi-year project funding through competitive bidding, with cumbersome and time-consuming procedures enhancing competition between laboratories. The recruitment is at its lowest, which is increasing drastically the precarity and competition. The implementation of the semi-automated academic and social selection system makes a clear barrier to access higher education equally. 
%LPR https://www.legifrance.gouv.fr/download/pdf?id=Ta4lC9NxVBJnpowWgmcZ8d_gRqcUA3qn9Cpuf_2cwJA=

There are still a few defenses against these attacks, such as organizations, unions, and political engagement. However, many colleagues have deserted them as indicated by the decreasing numbers of union membership \citep[in France, the number has fallen by a factor of four in sixty years,][]{insee2006syndicats} or the shift in volunteering and challenges facing the nonprofit sector  \citep{insee2006assoc, hely2014metamorphoses}. 

%%-------------------------
\section{Discussion \& conclusion}
In the discussion following the presentations, the conversation quickly turned to the debate on air versus rail travel for business trips. Although this is an easy lever to pull, in view of the results presented, it is by no means the major problem facing our community. Fixating on this might lead us in circles. We then shifted the conversation to exploring the community's potential commitment to making significant changes.  

One positive point is that the community has reached a fairly high level of awareness. Conversely, junior researchers feel that there's been little progress and that recent years' discussions have come up empty. This is fair. Yet, over the past five years, there's been a considerable shift in the acceptance of discussing these issues for the community's interest. This time frame seems slow compared to the faster evolution of our societies. 

Ultimately, we concluded that these discussions are eminently political, as reflected in the adage, "ecology without politics is just gardening," and that our community should engage in more political actions. Academic freedom is a democratic issue, as demonstrated by collective such as the organizations CAALAP\footnote{\url{https://caalap.fr/}} and ALIA\footnote{\url{https://liberte-academique.fr/}}. 
%Science sous l'occupation

\begin{acknowledgements}
The members of the SOC of the session and the members of the SF2A special commission \emph{Environmental transition} would like to warmly thank the SF2A council for supporting their activities, notably through the organization of the special afternoon session. 
D.R would like to thank Arnaud Saint-Martin, Jérôme Lamy and Pascal Marichalar for the many inspirations to his work. 
\end{acknowledgements}

\bibliographystyle{aa}  % A&A bibliography style file (aa.bst)
\bibliography{Cantalloube_S13} % your references in file: Yourfile.bib

\begin{thebibliography}{19}
\expandafter\ifx\csname natexlab\endcsname\relax\def\natexlab#1{#1}\fi

\bibitem[{Amossé \& Pignon(2006)}]{insee2006syndicats}
Amossé, T. \& Pignon, M.-T. 2006, Données sociales : La société française,
  4

\bibitem[{Barret {et~al.}(2024)Barret, Albouys, Kn{\"o}dlseder, Loizillon,
  d’Andrea, Ardellier, Bandler, Dieleman, Duband, Dubbeldam,
  {et~al.}}]{barret2024}
Barret, D., Albouys, V., Kn{\"o}dlseder, J., {et~al.} 2024, Experimental
  Astronomy, 57, 19

\bibitem[{Carlotto(2022)}]{carlotto2022bolsonaro}
Carlotto, M.~C. 2022, La Revue Nouvelle, 7, 29

\bibitem[{Chambaz {et~al.}(2021)Chambaz, Dezoteux, \& Redon}]{Chambaz2021}
Chambaz, B., Dezoteux, B., \& Redon, D. 2021, Le Centre spatial guyanais,
  Travers{\'e}e culturelle dans les archives de l'Espace \#2, ed. G.~Azoulay
  (CNES)

\bibitem[{Dablander {et~al.}(2024)Dablander, Sachisthal, \&
  Haslbeck}]{dablander2024}
Dablander, F., Sachisthal, M.~S., \& Haslbeck, J.~M. 2024, npj Climate Action,
  3, 105

\bibitem[{Dassas {et~al.}(2025)Dassas, Bonamy, Bzeznik, David, Frenoux,
  Guennebaud, Ligozat, Mallarino, \& Orgerie}]{dassas2025}
Dassas, K., Bonamy, C., Bzeznik, B., {et~al.} 2025, PhD thesis, EcoInfo

\bibitem[{dos Santos~Ilha {et~al.}(2024)dos Santos~Ilha, Boix, Kn{\"o}dlseder,
  Garnier, Montastruc, Jean, Pareschi, Steiner, \& Toussenel}]{dossantos2024}
dos Santos~Ilha, G., Boix, M., Kn{\"o}dlseder, J., {et~al.} 2024, Nature
  Astronomy, 8, 1468

\bibitem[{Dupret(2016)}]{dupret2016milei}
Dupret, X. 2016, La Revue Nouvelle, 1, 15

\bibitem[{{Fargette} {et~al.}(2025){Fargette}, {Devinat}, {Jarry}, {Le~Liboux},
  {Resseguier}, \& {Taupin}}]{Fargette2025}
{Fargette}, N., {Devinat}, M., {Jarry}, M., {et~al.} 2025, in SF2A-2025:
  Proceedings of the Annual meeting of the French Society of Astronomy and
  Astrophysics, ed. A.~{Siebert}, K.~{Bailli{\'{e}}}, M.~{B{\'{e}}thermin},
  F.~{Cantalloube}, E.~{Josselin}, N.~{Lagarde}, J.~{Malzac}, J.~{Richard},
  O.~{Selliez}, \& O.~{Venot}

\bibitem[{Ferdinand(2019)}]{Ferdinand2019}
Ferdinand, M. 2019, Une écologie décoloniale. Penser l’écologie depuis le
  monde caribéen (Seuil)

\bibitem[{Gardner {et~al.}(2021)Gardner, Thierry, Rowlandson, \&
  Steinberger}]{gardner2021}
Gardner, C.~J., Thierry, A., Rowlandson, W., \& Steinberger, J.~K. 2021,
  Frontiers in Sustainability, 2, 679019

\bibitem[{Goldman \& Chenoweth(2025)}]{goldman2025}
Goldman, G. \& Chenoweth, E. 2025, Scientists’ role in defending democracy

\bibitem[{Harbuz {et~al.}(2024)Harbuz, Dovgyi, Stryzhak, Ryabchyi, Prykhodniuk,
  Gorborukov, Nadutenko, Kurbatov, Taranov, Petrenko, {et~al.}}]{unesco2024}
Harbuz, T., Dovgyi, S., Stryzhak, O., {et~al.} 2024, Analysis of war damage to
  the Ukrainian science sector and its consequences (Paris, UNESCO)

\bibitem[{H{\'e}ly(2014)}]{hely2014metamorphoses}
H{\'e}ly, M. 2014, Les m{\'e}tamorphoses du monde associatif (Puf)

\bibitem[{Kn{\"o}dlseder(2025)}]{knodlseder2025}
Kn{\"o}dlseder, J. 2025, in SF2A-2025: Proceedings of the Annual meeting of the
  French Society of Astronomy and Astrophysics, ed. A.~{Siebert},
  K.~{Bailli{\'{e}}}, M.~{B{\'{e}}thermin}, F.~{Cantalloube}, E.~{Josselin},
  N.~{Lagarde}, J.~{Malzac}, J.~{Richard}, O.~{Selliez}, \& O.~{Venot}

\bibitem[{Kn{\"o}dlseder {et~al.}(2024)Kn{\"o}dlseder, Coriat, Garnier, \&
  Hughes}]{knodlseder2024}
Kn{\"o}dlseder, J., Coriat, M., Garnier, P., \& Hughes, A. 2024, Nature
  Astronomy, 8, 1478

\bibitem[{Marc {et~al.}(2024)Marc, Barret, Biancamaria, Dassas, Firmin,
  Gandois, Gheusi, Kuppel, Maisonobe, Mialon, {et~al.}}]{marc2024omp}
Marc, O., Barret, M., Biancamaria, S., {et~al.} 2024, PLOS Sustainability and
  Transformation, 3, e0000135

\bibitem[{Mariette {et~al.}(2022)Mariette, Blanchard, Bern{\'e}, Aumont,
  Carrey, Ligozat, Lellouch, Roche, Guennebaud, Thanwerdas,
  {et~al.}}]{mariette2022}
Mariette, J., Blanchard, O., Bern{\'e}, O., {et~al.} 2022, Environmental
  Research: Infrastructure and Sustainability, 2, 035008

\bibitem[{Prouteau \& Wolff(2013)}]{insee2006assoc}
Prouteau, L. \& Wolff, F.-C. 2013, Économie et Statistique, 459

\end{thebibliography}

%
%Hyperlinks can be introduced as follows: \url{http://www.sf2a.eu/}.
%Please use files in the PDF format, as shown in Fig.~\ref{author1:fig1}
%
\end{document}